\def\selectedoptions{final}

\documentclass[
   \selectedoptions
  ]
  {aipproc}

\layoutstyle{6x9}
\SetInternalRegister\hbadness{8000}
\newcommand\doingARLO[2][]{%
  \ifx\mmref\undefined #1\else #2\fi
}

\begin{document}

\title 
{On the formation of energy spectra of synchrotron X-rays and inverse Compton $\gamma$-rays in binary systems with luminous optical stars}


\author{D. Khangulyan}{
  address={MPI f\"ur Kernphysik, 69117 Heidelberg, Germany},
  email={dmitry.khangulyan@mpi-hd.mpg.de}
}

\author{F. Aharonian}{
  address={MPI f\"ur Kernphysik, 69117 Heidelberg, Germany},
  email={felix.aharonian@mpi-hd.mpg.de}
}

\copyrightyear  {2004}

\begin{abstract}
In this paper we study formation broad band energy spectra of electrons due to synchrotron radiation and inverse Compton (IC) scattering. The effect of transition of IC cooling from the Thomson regime to the Klein-Nishina regime can be very significant in the environments where radiation density dominates over the magnetic field density. We discuss impact of this effect on the formation of synchrotron and IC radiation components of TeV electrons in binary systems with very luminous optical stars. 
Such a scenario can be realized, for example, at the wind termination shock of the pulsar in PSR B1259-63/SS2883 or in the jets of microquasars. The calculations are based on the solution of Boltzmann equation using precise differential Compton cross-section $\left({\rm d}\sigma/ {\rm d}\Omega {\rm d}E\right)$.
\end{abstract}

\date{\today}

\maketitle
\section{Introduction}
In astrophysical objects where electrons are accelerated to very high energies at the presence of intense photon fields, the density of which significantly exceeds the density of magnetic field, the effect of transition of Compton cooling from the Thomson regime to Klein-Nishina regime plays important role in formation of electron spectra and related with them energy spectra of synchrotron and inverse Compton radiation components. A possible example of such a source is the binary pulsar PSR B1259-63/SS2883 \cite{johnston}. In this system a millisecond pulsar is orbiting around the massive Be2 star with tempreture $T\simeq 2.3\cdot10^4 {\rm K}$ and luminosity \mbox{$L_{\rm star}\simeq3.3\cdot10^{37}{\rm erg/s}$}. This implies huge photon  densities, especially at epochs of the pulsar passage near the periastron, when the density of photons achieves \mbox{$ w_{\rm ph}={ L_{\rm star}/4\pi R^2c}\simeq 1\cdot\left( R/10^{13}{\rm cm}\right)^{-2}{\rm erg/cm^3}$}. Thus, for the magnatic field $B\simeq1 {\rm G}$ (the corresponding energy destity $B^2/8\pi\simeq0.5{\rm erg/cm^3}$), which is estimated from the MHD treatment of the termination of the pulsar wind \cite{tavani97,ball}, the energy density of photon field exceeds the energy density of magnetic field. This means that while for electrons of relatively low energy the energy losses are dominated by inverse Compton scattering, for TeV electrons the losses are dominated by synchrotron radiation (because of the suppressed Compton cross-section due to the Klein-Nishina effect). Although  the cooling time in both energy regimes behaves as $t_{\rm cooling}\propto 1/E_{\rm e}$, the absolute scales differ by a factor of $3$. On the other hand in the transition rigion (from $10\ {\rm GeV}$ to $1\ {\rm TeV}$) the cooling time essentially deviate from the dependence $1/E_{\rm e}$. This leads to the formation of unusual spectra of electrons and the related electromagnetic radiation. Obviously this effect is very important for interpretation of synchrotron X-rays and IC $\gamma$-rays. Such radiation componets were predicted from the unique binary pulsar PSR B1259-63/SS2883 \cite{tavani97,kirk,murata}. 
A similar scenario can be realized in the microquasar LS 5039 although in this case the TeV electrons have a different origin -- most likely they are accelerated in the continuous jet originating from the compact star \cite{valenti}.
\begin{figure}
{\includegraphics[height=.45\textheight,angle=270.0]{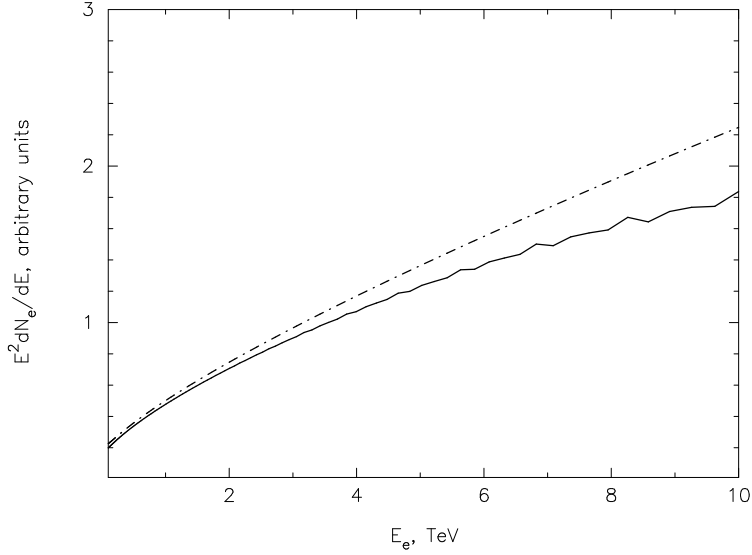}}
\caption{The steady state  energy distribution function of electrons obtained within the continuous lose approximation (dashed line) and from the exact solution of the Boltzmann equation (solid line). The temperature of blackbody radiation is assumed $T=2.3\cdot10^4$. The differential injection spectrum of electrons is taken as $q(E_{\rm e})\propto E_{\rm e}^{-2}$. The results are shown in arbitrary units. Note that the ignorance of the Klein-Nishina effect would give much steeper steady state distribution $E_{\rm e}^{2}{\rm d}N/{\rm d}E\propto E_{\rm e}^{-1}$}\label{losses}
\end{figure}

Bellow we investigate the importance of several effects for calculations of the electron distribution and corresponding synchrotron and IC energy spectra in binary systems with powerful optical star. These are, in particular, the approach of the continuous energy lose aproxiamtion for electrons, the Klein-Nishina effect and the anisotropy in distribution of the target radiation field.

\section{The electron distribution function}
To be specific, here we consider a systems similar to PSR B1259-63/SS2883 (for details see \cite{johnston,tavani97}). We performed the analysis of the system assuming a power-law injection spectrum for electrons with exponential cut-off formed at the termination shock \cite{kennel}. The energy distribution of electrons is determined by radiative losses; the impact of adiabatic losses we discuss elsewhere. 
\begin{figure}
{\includegraphics[height=.45\textheight,angle=270.0]{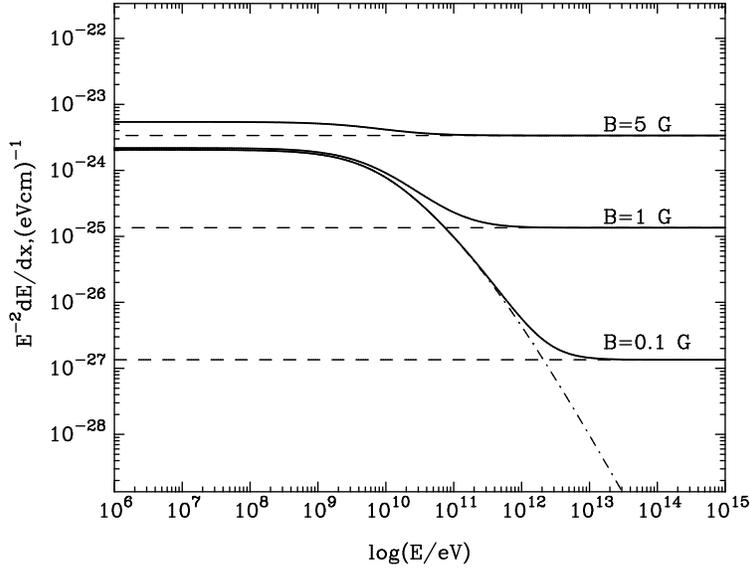}}
\caption{Energy lose rates of electrons. The dashed lines correspond to the synchrotron losses, the dot-dashed lines to IC losses, and the solid lines to the total radiative losses.}\label{eloss}
\end{figure}

Although the continous energy lose aproximation is often used for determination of the electron distributions in ambient radiation filed, this approximation \textit{a priory}  does not guarantee adequate  accuracies, especially in the Klein-Nishina regime where the interactions have catastrophical character. The applicability of this approach we inspected comparing the results with ones obtained from the solution of the Boltzmann equation which appropriately treats the catastrophic energy losses (see Fig.\ref{losses}). In order to demonstrate the difference of the results, we assumed that the magnetic field $B=0$. One can see that, despite the fact that the scattering proceeds in the deep Klein-Nishina regime (the parameter \mbox{$b= 4(3kT)E_{\rm e}/(mc^2)^2\simeq90(E_{\rm e}/TeV)\gg1$}), the continuous lose approximation describes the electron distribution within an accuracy less than $20\%$ for $E_{\rm e}\leq10\ {\rm TeV}$. Since this energy range is relevant for the production of synchrotron X-rays and TeV $\gamma$-rays, in what follows we present results obtained within much simpler continuous energy lose approximation.
\begin{figure}
{\includegraphics[height=.45\textheight,angle=270.0]{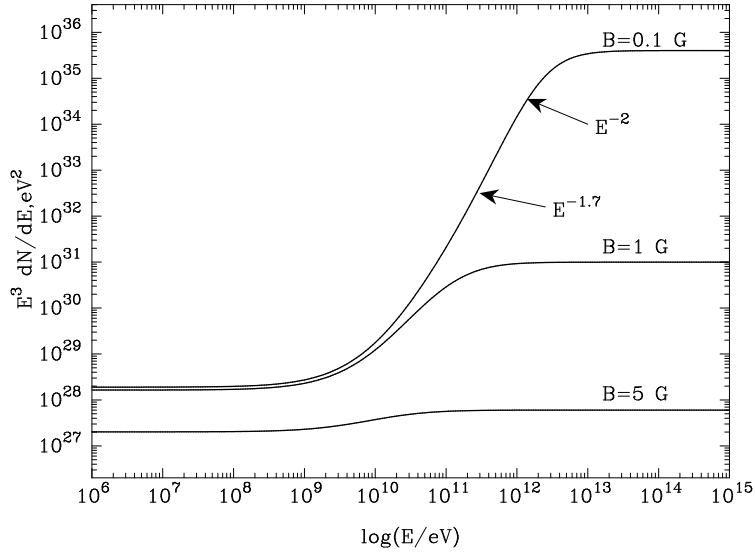}}
\caption{The electron steady state energy distribution functions for three different magnatic fields. The electron injection spectrum is assumed in the form $E_{\rm e}^{-2}e^{-E_{\rm e}/E_{\rm cut}}$ with $E_{\rm cut}\rightarrow \infty$.}\label{density}
\end{figure}
\begin{figure}
{\includegraphics[height=.4\textheight,angle=270.0]{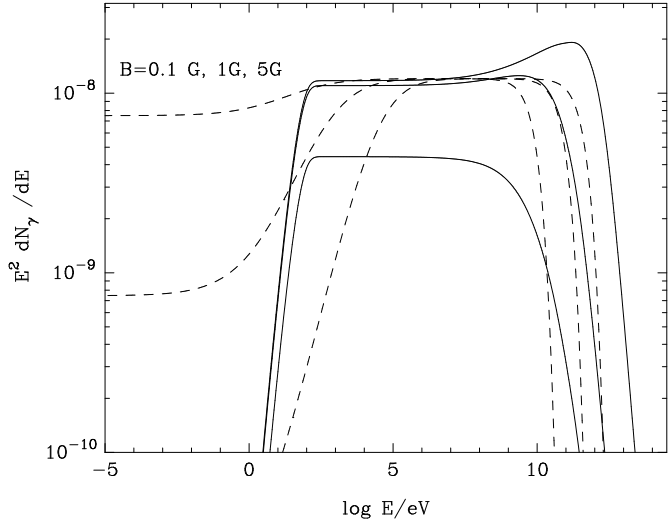}}
\caption{Spectral energy distributions of synchrotron radiation (dashed lines) and IC $\gamma$-rays (solid lines) calculated for the steady state electron distributions shown in Fig.\ref{density}}\label{rad0}
\end{figure}

Within this approach the steady state distribution of electron has the following form $N_e(E_{\rm e})={1\over|\dot{E}_{\rm e}|}\int_{E_{\rm e}}^\infty q(E_{\rm e}){\rm d}E_{\rm e}$:
here $q(E_{\rm e})$ is injection rate and $\dot{E}_{\rm e}$ is energy lose rate. The synchrotron and IC losses in the Thomson regime have the same energy dependence $\dot{E}_{\rm e}\propto E_{\rm e}^2$; in this case the ratio of synchrotron and IC losses equals to the ratio of the energy densities in the magnetic and photon fields. 
In the Klein-Nishina regime the Compton lose rate is almost energy independent $\dot{E}_{\,ic}\propto const$. Thus, in the radiation dominated environments 
energy lose rate may have very specific energy dependence.

This is demonstrated in  Fig.\ref{eloss} for a fixed value of the photon energy density $w=1 \ {\rm erg/cm^3}$ and temperature $T=2.3 \cdot 10^4 {\rm K}$, but for three different values of the
ambient magnetic field, $B=0.1$ G, 1 G, and  5 G. 
\begin{figure}
{\includegraphics[height=.7\textheight,angle=270.0]{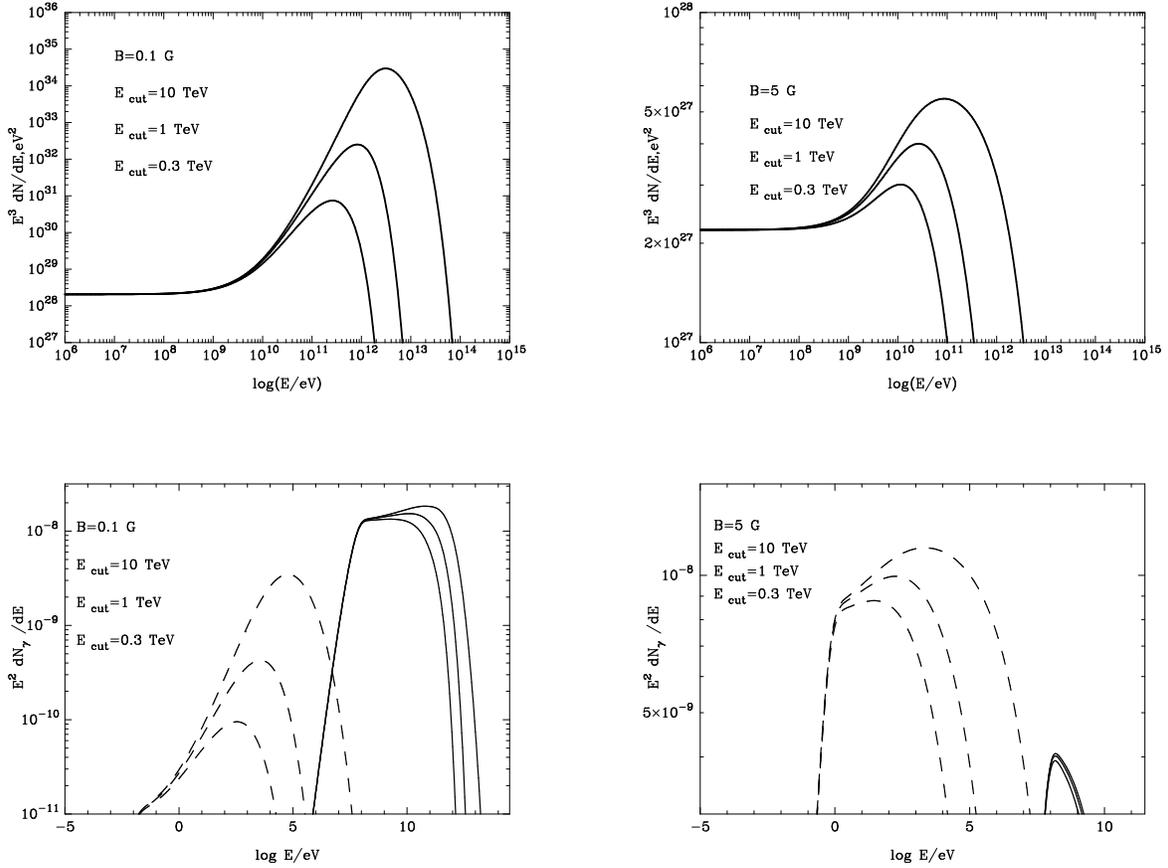}}
\caption{Top panel: The same as in Fig.\ref{density} but for fixed magnetic fields, $B=0.1$G (left panel) and $B=5$G (right panel) and for three different cutoff energies, $E_{\rm cut}=0.3$ TeV, 1 TeV and 10 TeV. Bottom panel: Spectral energy distributions of synchrotron (dashed lines) and IC (solid lines) radiation components calculated for the steady state electron distributions shown in the top panel.}\label{rad}
\end{figure}
Correspondingly, the steady state spectra of electrons significantly depend 
on the strength of the magnetic field. While the synchrotron and Thomson losses 
make the electron spectrum steeper, the losses in the Klein-Nishina regime
at the presense of a low magnetic field lead to the hardening of the electron spectrum. In the case when  the magnetic field density is comparable or less than the target photon density, one should expect quite irregular electron spectra, especially in the transition region (see Fig.\ref{density}). The electron spectra shown in this figure are calculated for a pure power-law injection spectrum, $q(E_{\rm e})\propto E_{\rm e}^{-2}$, for three different magnatic fields. In the case of losses dominated by synchrotron radiation and/or by Thomson scattrering, the steady state spectrum of electrons is ${\rm d}N/{\rm d}E\propto E_{\rm e}^{-3}$. On the other hand, at the absence of magnatic field the steady state electron spectrum formed in the deep Klein-Nishina regime is very hard, close to ${\rm d}N/{\rm d}E\propto E_{\rm e}^{-1}$. Howether, since for the assumed magnetic fields and the energy density of the star light, the broad band spectra of electrons have more complicated forms as shown in Fig.\ref{density}. In low energy region we see indeed $E_{\rm e}^{-3}$ type spectrum with significant hardening in the transition region from $10\ {\rm GeV}$ to $1\ {\rm TeV}$. Note that at very high energies the Compton losses are suppressed and the synchrotron losses dominate; therefore the electron spectrum  
again obtains a $E_{\rm e}^{-3}$ form. However, in reality the spectrum at such high energies depends strongly on the high-energy cutoff in the injection spectrum. This is demonstrated in Fig.\ref{rad} (top panel), where the injection spectrum is given in the form 
$q(E_{\rm e})=E_{\rm e}^{-2}e^{-E_{\rm e}/E_{\rm cut}}\,$, with three different cutoff energies, $E_{\rm cut}=0.3$, 1, and 10\, \rm TeV. 

\section{Spectral energy distribution of radiation}
The spectra of synchrotron and IC radiation components corresponding to the electron distributions in Fig.\ref{density} are shown in Fig.\ref{rad0}. Note that in this figure the synchrotron spectra extend to very high energies, which is the result of the assumption of absence of the high energy cutoff in the injection spectrum of electrons. On the other hand the IC spectrum extends to very low energies because of absence of a low-energy cutoff in the injection spectrum of electrons. In Fig.\ref{rad} we present more realistic spectra assuming both low- and high-energy cutoffs in the electron injection spectrum. In this case the synchrotron and IC components 
are well separated in the spectral energy distribution of radiation.

\section{The impact of the anisotropy of the target photon field}
\begin{figure}
{\includegraphics[height=.65\textheight,angle=270.0]{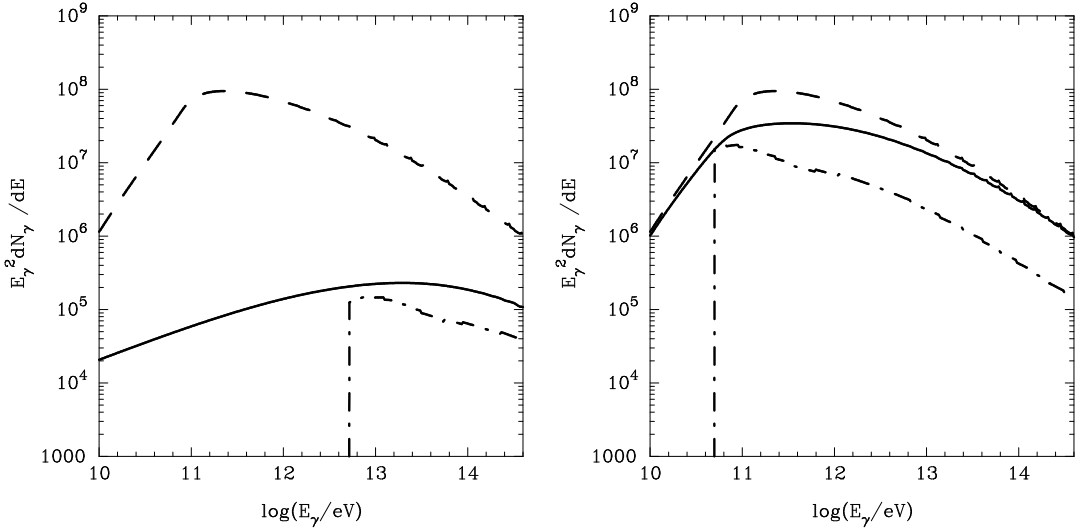}}
\caption{Spectral energy distribution of the IC radiation for a truncated power-law electron spectrum, \mbox{$E_{\rm e}^{-2}( 100{\rm GeV}\leq E_{\rm e}\leq10^3{\rm TeV})$}, calculated using the accurate cross-section (solid lines), the cross-section average over the angle (dashed lines) and $\delta$-function approximation of the cross-sectiont (dot-dashed lines), calculated for two different interaction angles, $\theta=5^{\circ}$(left panel) and $\theta=45^{\circ}$(right panel).}\label{angl}
\end{figure}

Since in the binary system the relativistic electrons "see", for any given time, the radiation from the star at certain angles, it is important to use the precise double differential cross-section, ${\rm d}\sigma/{\rm d}\Omega {\rm d}E$, for calculations of both the spectral and temporal characteristics of radiation. This is demonstrated in Fig.\ref{angl} where the spectral energy distribution of the IC radiation is shown for different interaction angles using (i) the accurate cross-section, (ii) cross-section average over the angle, and (iii)  $\delta$-functional  approximation like the one 
used in Ref. \cite{kirk}. The results show that both the average cross-section 
and $\delta$-functional approximation fail to describe appropriately 
the broad band energy spectrum,  as well as absolute $\gamma$-ray fluxes. 

\hyphenation{Post-Script Sprin-ger}

\end{document}